\documentclass[hyper]{JHEP3}  
\usepackage{cancel} 
\usepackage{amssymb}     
\usepackage{amsmath}
\usepackage{graphicx,subfig} 
\usepackage{multirow} 
\usepackage{dsfont} 
\usepackage{cite}

\numberwithin{equation}{section} 
\numberwithin{figure}{section}
\numberwithin{table}{section}

\title{Conformal windows of $SP(2N)$ and $SO(N)$ gauge theories from topological excitations on $\mathds{R}^3\times S^1$} 
\author{Siavash Golkar\\
		Department of Physics, University of Toronto, Toronto, ON M5S-1A7, Canada\\
		E-mail: \email{sgolkar@physics.utoronto.ca}
		\thanks{Changing to golkar@u.washington.edu as of October 1$^\text{st}$ 2009.}}
\abstract{We derive an estimate of the lower boundary of the conformal window of  $SP(2N)$ and $SO(N)$ gauge theories with fermionic matter in several different representations. We calculate the index of topological excitations for these groups on the manifold $\mathds R^3\times S^1$, from which we deduce the scale of the generation of the mass gap of the theory. This is then used to approximate the critical value of the number of species $n_f^*$ for the onset of conformality on $\mathds R^4$. We also provide a detailed comparison with other estimates of the conformal window.}

\preprint

\begin{document}

\section{Introduction}

\subsection{Motivation}

In recent years, theories with strong gauge dynamics at or near conformality, specifically those in four dimensions, have been receiving an increasing amount of attention. One of the most pertinent areas of their applicability is arguably the problem of electroweak symmetry breaking. It is conceivable that the breaking of electroweak symmetry might be induced by a strongly coupled gauge theory without elementary Higgs scalars. It is well known that utilizing an appropriately rescaled QCD cannot explain electroweak precision data, however, it has been argued that gauge sectors with near-conformal or conformal behavior can help solve phenomenological problems of fine-tuning and flavor in models of dynamical electroweak symmetry breaking \cite{Holdom:81,Yamawaki:85,Appelquist:86,Luty:04}. The more recent formalism of ``unparticle physics'' is also heavily dependent on conformal theories \cite{Georgi:07}.

However, the very nature of these strongly coupled gauge theories makes them difficult subjects of study to the point that there is currently no definitive method of determining which asymptotically free gauge theories flow to conformal field theories in the infrared. Given the relevance of these theories in different areas of high energy physics, the exploration of the parameter space is surely warranted. 

\subsection{Outline}
The general idea behind our calculation is given in section \ref{sec:conf-bound}. The reader primarily interested in our final results can skip over to section \ref{sec:Comparisons} where our results have been summarized in tables \ref{tab:C_N-fund} through \ref{tab:SO(N)-sym}. In the rest of the paper we provide a more detailed calculation of the conformal window of several classes of gauge theories.

We begin in chapter \ref{sec:Index}, where we calculate the index of topological excitation on $\mathds R^3 \times S^1$ by generalizing the results of \cite{Poppitz:Index}. In section \ref{sec:Monopole-Backgrounds} we introduce the backgrounds of interest which are comprised of Prasad-Sommerfield solutions of $SU(2)$ embedded into a general gauge group $G$ as well as the four dimensional instanton and their combinations. In \ref{sec:Index-calculation} we explicitly derive the index of these objects and demonstrate that for small circle sizes, unsurprisingly, it is equal to the Callias index \cite{Callias:77}.

Section \ref{sec:conf-bound} is the heart of our discussion. After a brief overview of the methodology, we provide detailed calculations of the mass gap as well as the conformal windows of several different representations of $SP(2N)$, $SO(2N+1)$ and $SO(2N)$ gauge groups, respectively in sections \ref{sec:C_N}, \ref{sec:B_N} and \ref{sec:D_N}. These results are summarized in tables \ref{tab:C_N-fund} through \ref{tab:SO(N)-sym} in section \ref{sec:Comparisons}, where we also provide detailed comparisons with the results of two other approaches: the ladder approximation using truncated Schwinger-Dyson equations and the NSVZ-inspired approach with both $\gamma=1$ and $\gamma=2$, as well as the estimates from the worldline formalism\cite{Pagels:75,Appelquist:1988yc,Dietrich:2006cm,Sannino:08,Sannino:09,Armoni:09}.

In section \ref{sec:Conclusion} we conclude, discuss the shortcomings of our approach and provide possible directions for future work. Section \ref{sec:notation} provides explains our notation and section \ref{app:Lie_Prel} summarizes a few facts about Lie algebras we use throughout the paper.

\subsection{Notation}\label{sec:notation}

We work with the four dimensional Euclidean Dirac operator in representation $R$:
\begin{equation}
	\cancel D=\gamma_\mu D_\mu \quad ,\quad D_\mu=\partial_\mu+iA_\mu^a T^a,
	\label{eq:Dirac-op-defined}
\end{equation}
where the matrices $T^a$ (along with other Lie algebra related notations) are defined in appendix \ref{app:Lie_Prel}. Throughout the paper, Roman indices from the beginning of the alphabet are Lie algebra indices, Roman indices from the middle of the alphabet run from 1 to 3 and correspond to the non-compactified coordinates, Greek indices from the beginning of the alphabet refer to roots of the Lie algebra and Greek indices from the middle of the alphabet run from 1 to 4 and correspond to the four dimensional manifold. We also take $x^4\equiv y$ to be the direction along the compactified dimension with $y\sim y+L$.

Our choice of the gamma matrices is:
\begin{equation}
		\gamma_k=\sigma_1\otimes\sigma_k\;,\;\gamma_4=-\sigma_2\otimes\mathds 1\;,\;\gamma5=\sigma_3\otimes\mathds 1,
		\label{eq:gamma-defined}
\end{equation}
where $\sigma_k$ are the Pauli matrices. The vector-like Dirac operator is then:
\begin{equation}
		\cancel D=
			\begin{pmatrix}
					0																	&		\sigma_k D_k+i\sigma_0 D_4\\
					\sigma_k D_k - i \sigma_0 D_4 0		&		0
			\end{pmatrix}
		\equiv 
			\begin{pmatrix}
					0		&		-D^\dagger\\
					D		&		0
			\end{pmatrix},
		\label{eq:Dirac-expanded}
\end{equation}
where we have defined the Weyl operator $D$. We have:
\begin{subequations}
	\begin{align}
		D D^\dagger &=-D_\mu D^\mu + \sigma_m\frac12 \epsilon_{mkl}F_{kl}-\sigma_k F_{4k}=-D_\mu D^\mu,\\
		D^\dagger D &=-D_\mu D^\mu + \sigma_m\frac12 \epsilon_{mkl}F_{kl}+\sigma_k F_{4k}=-D_\mu D^\mu+2\sigma^m B^m,
	\end{align}
	\label{eq:D-properties}
\end{subequations}
where $D_\mu$ is as in \eqref{eq:Dirac-op-defined} and we have assumed, without loss of generality, that the background is anti-self-dual, i.e. $F_{4k}=\frac12 \epsilon_{klm}F_{lm}=B_k$.

\section{The Index of the Dirac operator}\label{sec:Index}

In this section we rewrite the results of \cite{Poppitz:Index} for general gauge groups\footnote{These results can be derived directly from \cite{Singer:00}.}. Following their notation we let $U(x)$ be the holonomy of the Wilson line wrapping around the compact direction:
\begin{equation}
	U=P\exp i \oint A_4 \text dy, \qquad x\in \mathds R^3, y\in S^1,
	\label{eq:holonomy}
\end{equation}
where $A_4$ is the component of the gauge field along the circle. In the weak coupling regime, $A_4$ behaves as a compact adjoint Higgs field and its asymptotic value $A_4|_\infty$ induces gauge symmetry breaking. Here, we assume that the symmetry breaking is maximal or equivalently $\alpha(A_4|_\infty)\neq 0,\, \forall \alpha\in\Phi$. This allows us to choose a base $\Delta$, with respect to which $A_4$ is strictly dominant.

\subsection{Monopole Backgrounds}\label{sec:Monopole-Backgrounds}

There are two classes of nontrivial topological excitations on $\mathds R^3\times S^1$ \cite{Lee:98,Lee:97}. The first is the generalization of the three-dimensional $SU(2)$ Prasad-Sommerfield monopole and is $y$-independent. For a gauge group $G$ of rank $l$, there are $l$ independent such solutions obtained by embeddings of the $SU(2)$ solution in $G$. We construct them as follows.

Let $\varphi^s(x,\lambda)$ and $A_j^s(x,\lambda)$ be the $SU(2)$ monopole solution corresponding to a vacuum expectation value $\lambda$. In regular (hedgehog) gauge we have:
\begin{equation}
\begin{split}
	\varphi^s(x,\lambda)&=\frac{\hat x^s}{|x|}\left(1-\lambda|x|\coth \lambda|x|\right),\\
	A_j^s(x,\lambda)&=\epsilon_{sjk}\hat x^k\left(\frac1{|x|}-\frac{\lambda}{\sinh \lambda |x|}\right).
\end{split}
\label{eq:SU2-Solution}
\end{equation}
 Let $\alpha$ be any root and take:
\begin{equation}
\begin{split}
t^1_\alpha&=\frac{1}{2}(x_\alpha+y_\alpha),\\
t^2_\alpha&=\frac{i}{2}(y_\alpha-x_\alpha),\\
t^3_\alpha&=\frac{1}{2}h_\alpha,
\end{split}
\label{eq:SU(2)_embed}
\end{equation}
with $x_\alpha,\; y_\alpha$ and $h_\alpha$ as in appendix \ref{app:Lie_Prel}. Then,
\begin{equation}
	\begin{split}
			A_4(x)&=\sum_{s=1}^3\varphi^s(x,\lambda)t_\alpha^s+A_4|_\infty-\alpha(A_4|_\infty) t^3_\alpha,\\
			A_j(x)&=\sum_{s=1}^3A_j^s(x,\lambda)t_\alpha^s,
	\end{split}
	\label{eq:Monopole_Embed}
\end{equation}
with $\lambda=-\alpha(A_4|_\infty)$, is an elementary anti-self-dual monopole solution corresponding to the root $\alpha$, satisfying $F_{4k}=-\frac12\epsilon_{4kpq}F_{pq}=\frac12\epsilon_{kpq}F_{pq}=B_k$ \cite{Weinberg:80}. The asymptotic behavior of this solution in the string gauge is given as:
\begin{equation}
\begin{split}
A_4&\to A_4|_\infty-\frac1{|x|} t^3_\alpha,\\
B_j&\to\frac{\hat x_j}{|x|^2} t^3_\alpha,
\end{split}
\label{eq:monopole-Asymptote}
\end{equation}
which gives magnetic charge $-\alpha^*$ and topological charge $\frac{L}{2\pi}\alpha^*(A_4|_\infty)$. The independent monopoles are the solutions corresponding to the $l$ simple roots.
 
The second class is comprised of solutions that wrap around the compact dimension \cite{Lee:98,Lee:97}, most of these can be obtained from the first class solutions via non-periodic gauge transformations. We start with the solution \eqref{eq:Monopole_Embed} corresponding to a root $\alpha$ and we apply the gauge transformation:
\begin{equation}
U_\alpha(y)=\exp\left(\frac{i 2\pi}{L}y t^3_\alpha\right).
\label{eq:nonper_gauge_trans}
\end{equation}
However, the asymptotic value of this solution is shifted by:
\begin{equation}
\left.A'_4(x,y)\right|_{|x|\to\infty}=A^4_\infty-\frac{2\pi}{L} t^3_\alpha.
\label{eq:A_4-shift}
\end{equation}
To compensate for this shift, we use the solution \eqref{eq:Monopole_Embed} with a vacuum expectation $A_4|_\infty$ shifted in the other direction. The new solution is then given by:
\begin{equation}
\begin{split}
A&_4(x)=\sum_{s=1}^3\varphi^s(x,\lambda)t'^s+A_4|_\infty-\alpha(A_4|_\infty) t^3_\alpha-\frac{2\pi}{L}t^3_\alpha,\\
A&_j(x)=\sum_{s=1}^3A_j^s(x,\lambda)t'^s,
\end{split}
\label{eq:Twisted-Monopole}
\end{equation}
with $\lambda=-\alpha(A_4|_\infty+\frac{2\pi}{L} t^3_\alpha)$ and $t'^s$ defined as:
\begin{equation}
t'^s_\alpha=U_\alpha  t_\alpha^s U_\alpha^{-1}.
\label{eq:SU(2)-Conj}
\end{equation}
These transformations do not change the magnetic charge of the solution but decrease the topological charge by 1. One can therefore consider this new solution as a monopole corresponding to the root $\alpha$ combined with an (anti)-instanton of charge -1. In the same way, one can construct towers of solutions on top of the $l$ basic solutions \eqref{eq:Monopole_Embed} by consecutively applying the gauge transformation \eqref{eq:nonper_gauge_trans} while compensating the vacuum shift in a similar manner to the solution obtained above.

However, there exists an independent tower of solutions built on top of the lowest root $\tilde\alpha$, defined in \eqref{eq:alpha-tilde}. Note that although these solutions correspond to a negative root, they are not anti-monopoles which would satisfy self-dual equations (Remember we defined our monopoles to satisfy anti-self-dual equations.) \cite{Lee:98,Lee:97,Davies:00}. We construct them as follows. Start with the solution \eqref{eq:Monopole_Embed} for $\alpha=-\tilde\alpha$.

First, we define a new solution with an asymptotic value $A_4'|_\infty=\sigma_{\alpha}(A_4|_\infty)+n\pi h_{\alpha}$, where $\sigma_\alpha$ is the Weyl reflection
\footnote{We use the correspondence between the elements of the Cartan and the weights explained in appendix \ref{app:Lie_Prel} to define $\sigma_{\alpha}(A_4|_\infty)$. If $w$ is the weight corresponding to $A_4|_\infty$ then
\begin{equation}
\sigma_\alpha(A_4|_\infty)=t_{\sigma_{\alpha(w)}}.
\label{A_4-Reflection}
\end{equation}}
 in $\alpha$. Then we perform a Weyl reflection and the gauge transformation \eqref{eq:nonper_gauge_trans} to restore the asymptotic value of $A_4$. The final solution is:
\begin{equation}
\begin{split}
A&_4(x)=\sum_{s=1}^3\varphi^s(x,\lambda)t'^s+A_4|_\infty + \alpha(A_4|_\infty) t^3_\alpha-\frac\pi{L}h_{\alpha},\\
A&_j(x)=\sum_{s=1}^3A_j^s(x,\lambda)t'^s,
\end{split}
\label{eq:affine-monopole}
\end{equation}
with $\lambda=-\alpha(A_4'|_\infty)$ and $t'^s$ defined as:
\begin{equation}
t'^s=U_{\alpha}  \sigma_{\alpha}(t_\alpha^s) U_{\alpha}^{-1}.
\label{eq:SU(2)-Conj-Refl}
\end{equation}
The magnetic charge of this solution is $-\tilde\alpha^*$ and the topological charge is $\frac{L}{2\pi}\tilde\alpha^*(A_4|_\infty)-1$. 
\subsection{Calculating the index}\label{sec:Index-calculation}

	Now, we calculate the number of zero modes associated with the excitations discussed above. We follow the procedure of \cite{Poppitz:Index}. The Callias index of a Weyl fermion  for a representation $R$ is defined as:
\begin{equation}
	I_R=\lim_{M^2\to0} \text{Tr}\frac{M^2}{D^\dagger D+M^2}-\text{Tr}\frac{M^2}{D D^\dagger+M^2}.
	\label{eq:Callias-Weyl}
\end{equation}
This function counts the difference of the number of zero modes of the operator $D$ and the operator $D^\dagger$. We rewrite this as:
\begin{equation}
	I_R(M^2)=\text{Tr}\gamma_5\frac{M^2}{-\cancel D^2+M^2}=M\text{Tr}\gamma_5\frac{\cancel D + M}{-\cancel D^2+M^2}
	=-M\text{Tr}\gamma_5\frac{1}{\cancel D-M},
	\label{eq:Callias-Dirac}
\end{equation}
where in the second equality holds because of the cyclicity of the trace. We note that the expression in the trace is just the propagator of a fermionic field with action $S=\bar \psi\left(\cancel D-M\right)\psi$. For such theories in a locally four dimensional background we have:
\begin{equation}
	\partial_\mu J^5_\mu=\partial_\mu\left(\bar\psi\gamma_\mu\gamma_5\psi\right)
		=-2M\bar\psi\gamma_5\psi-\frac1{8\pi^2}T(R)\epsilon_{\mu\nu\rho\sigma}G^a_{\mu\nu}G^a_{\rho\sigma},
	\label{eq:J^5-identity}
\end{equation}
where the first term is the explicit axial symmetry breaking and the second term is the topological contribution to the axial current. Using this identity we rewrite \eqref{eq:Callias-Dirac} as:
\begin{equation}
	\begin{split}
		I_R(M^2)&=-M\text{Tr}\gamma_5\left<\psi\bar\psi\right>
				=M\int\text d^3x \int\limits_0^L \text dy \left<\bar\psi\gamma_5\psi\right>,\\
			&=-\frac12\int\limits_{S^2_\infty} \text d^2\sigma^k\int\limits_0^L\text dy\left<J_k^5\right>
				-\frac{T(R)}{16\pi^2}\int\text d^3x \int\limits_0^L\text dy 
				\epsilon_{\mu\nu\rho\sigma}G^a_{\mu\nu}G^a_{\rho\sigma},
	\end{split}
	\label{eq:Index-Axia-Current}
\end{equation}
where in the third equality the surface integral of $\partial_y\left<J_4^5\right>$ over the compact direction vanishes. Having separated the contribution of the topological charge, we now revert the first term to a form similar to \eqref{eq:Callias-Weyl} by doing the manipulations that took us from \eqref{eq:Callias-Weyl} to \eqref{eq:Callias-Dirac} in the other direction. Using the explicit form of the operators from \eqref{eq:Dirac-expanded} and \eqref{eq:D-properties}, we get:
\begin{equation}
	\begin{split}
		I_R^1(M^2)&\equiv-\frac12\int\limits_{S^2_\infty} \text d^2\sigma^k\int\limits_0^L\text dy\left<J_k^5\right>
				=-\frac12\int\limits_{S^2_\infty} \text d^2\sigma^k\int\limits_0^L\text dy\left<x\right|\gamma^k\gamma_5\cancel D
					\frac1{-\cancel D^2+M^2}\left|x\right>\\
			&=\frac12 \int_{S^2_\infty}{ \text d^2\sigma^k \int_0^L{ \text dy \, \text{tr}\,
				\left<x\right|\sigma^k\sigma^lD_l\left(\frac1{-D_\nu^2+M^2+2\sigma^mB^m}
				-\frac1{-D_\nu^2+M^2}\right)\left|x\right>}}\\
			&\quad-\frac12 \int_{S^2_\infty} \text d^2\sigma^k \int_0^L \text dy \, \text{tr}\,
				\left<x\right|i\sigma^kD_4\left(\frac1{-D_\nu^2+M^2+2\sigma^mB^m}
				+\frac1{-D_\nu^2+M^2}\right)\left|x\right>.
	\end{split}
	\label{eq:I^1-Final}
\end{equation}
This brings us to the final form of the index formula which will be used for the calculation:
\begin{equation}
		I_R(M^2)=I^1_R(M^2)-2T(R)Q,
		\label{eq:Index-final}
\end{equation}
where $I^1_R$ is defined above and $Q=\dfrac{1}{32\pi^2}\int\text d^3x \int_0^L\text dy\epsilon_{\mu\nu\rho\sigma}G^a_{\mu\nu}G^a_{\rho\sigma}$ is the topological charge, the values of which are given in section \ref{sec:Monopole-Backgrounds}. 

We now compute the contribution of the surface term $I^1_R$ for a composite monopole made up of $n_\alpha$ fundamental monopoles of type $\alpha$, where $\alpha \in \tilde \Delta$.

The integrals in \eqref{eq:I^1-Final} are evaluated at infinity in $\mathds R^3$ where we have:
\begin{equation}
		-D_\nu^2+M^2\approx -\partial ^2_m+M^2-D_4^2, \text{ with } -iD_4\to\frac{2\pi n}{L}+A_4.
		\label{eq:D-Asymptote}
\end{equation}
Using this and \eqref{eq:monopole-Asymptote}, we expand \eqref{eq:I^1-Final} in $B^k$. The first term contribution vanishes after taking the Pauli matrix trace and we have:
\begin{equation}
		\begin{split}
				I^1_R(M^2)&=2\int\limits_0^L\text dy\int\limits_{S^2_\infty} \text d^2\sigma^k \text {tr}\left<x,y\right|iD_4\frac{1}{-\partial_m^2+M^2-D_4^2}B^k
										\frac{1}{-\partial_m^2+M^2-D_4^2}\left|x,y\right>,\\
						&=-2\int\limits_0^L\text dy\int\limits_{S^2_\infty} \text d^2\sigma^k \text {tr}\left<x,y\right|(\frac{2\pi n}{L}+A_4)^2
								\frac{1}{\left[-\partial_m^2+M^2+(\frac{2\pi n}{L}+A_4)^2\right]^2}\frac{\hat x^k}{|x|^2}t^3_\alpha\left|x,y\right>.
		\end{split}
		\label{eq:expanding-I_1}
\end{equation}
Evaluating these integrals ($\text d^2\sigma^k=|x|^2x^k\text d\Omega_{S^2}$) at the $M\to 0$ limit, the surface contribution to the index of this excitation is:
\begin{equation}
		I^1_R(0)= -\sum_{\alpha\in\tilde\Delta}\sum_{n\in\mathds Z}n_\alpha\text{Tr}_R \left[\text{sign}\left(\frac{2\pi n}{L}+A_4|_\infty\right)\, t^3_\alpha\right].
		\label{eq:I^1-unregulated}
\end{equation}
This sum is divergent. We regulate it following \cite{Poppitz:Index}:
\begin{equation}
		\sum_{n=-\infty}^\infty \text{sign}(x+n)\to 1-2\left(x-\left\lfloor x\right\rfloor\right),
		\label{eq:regulator}
\end{equation}
where $\left\lfloor x\right\rfloor= \text {Max}\left\{n\in\mathds Z|n\leq x\right\}$ is the floor function. We have: 
\begin{equation}
	\begin{split}
			I^1_R(0)&=2\sum_{\alpha\in\tilde\Delta}n_\alpha\text{Tr}_R\left[\left(\frac{L}{2\pi}A_4|_\infty
												-\left\lfloor \frac{L}{2\pi}A_4|_\infty\right\rfloor\right) t^3_\alpha \right]\\
					&=T(R)\sum_{\alpha\in\tilde\Delta}n_\alpha\alpha^*\left(\frac{L}{2\pi}A_4|_\infty\right)
												-2\sum_{\alpha\in\tilde\Delta}n_\alpha\text{Tr}_R\left[\left\lfloor \frac{L}{2\pi}A_4|_\infty\right\rfloor t^3_\alpha\right],
	\end{split}
	\label{eq:I^1-regulated}
\end{equation}
where we have used the tracelessness of $h_\alpha$.
Plugging this into \eqref{eq:Index-final} and using the values of the topological charge from section \ref{sec:Monopole-Backgrounds} we have:
\begin{equation}
	\begin{split}
			I_R&= T(R)\sum_{\alpha\in\tilde\Delta}n_\alpha\alpha^*\left(\frac{L}{2\pi}A_4|_\infty\right)
												-2\sum_{\alpha\in\tilde\Delta}n_\alpha\text{Tr}_R\left[\left\lfloor \frac{L}{2\pi}A_4|_\infty\right\rfloor t^3_\alpha\right]\\
					&\qquad-2T(R)\sum_{\alpha\in\tilde\Delta}n_\alpha\alpha^*(\frac{L}{2\pi}A_4|_\infty) + 2n_{\tilde\alpha}T(R)\\
					&=2 T(R) n_{\tilde\alpha} - 2\sum_{\alpha\in\tilde\Delta}n_\alpha\text{Tr}_R\left[\left\lfloor \frac{L}{2\pi}A_4|_\infty\right\rfloor t^3_\alpha\right].
	\end{split}
	\label{eq:Index-final-regulated}
\end{equation}
We see that the non-integer contributions cancel and the final result is an integer.

From equation  \eqref{eq:Index-final-regulated} we can see that if we take the $n_\alpha$ such that $\tilde n=1$ and   $\tilde\alpha^*+\sum n_\alpha\alpha^*=0$, then the composite object would have zero magnetic charge, topological charge equal to -1 and index  equal to $2T(R)$. These are the quantum numbers of the four dimensional (anti-)instanton. We can therefore conclude that the $\mathds R^4$ instanton is made up of monopoles with multiplicities  given by the coefficients $n_\alpha$, called the Kac labels, given in table \ref{tab:Lie-data}. This also shows that for  the semi-classical expansion, composite objects made up of a few monopoles are more relevant than instantons.

For future reference, we also calculate the index in the small $L$ limit, where equation \eqref{eq:Index-final-regulated} reduces to:
\begin{equation}
	I_R=2T(R)n_{\tilde\alpha}-\sum_{\alpha\in\tilde\Delta} n_\alpha \text{Tr}_R\left[\text{sign}\left(\frac{L}{2\pi}A_4|_\infty\right)t^3_\alpha\right],
	\label{eq:Index-small-L}
\end{equation}
which is nothing but the Callias index on $\mathds R^3$ \cite{Callias:77}. 

\section{Conformality bounds}\label{sec:conf-bound}

	In this section we calculate approximate bounds for the onset of conformality for QCD-like theories with general gauge groups. We follow the procedures developed in \cite{Poppitz:Conf}, which we summarize here.

The main idea behind our analysis is that, with a few assumptions,  some properties of gauge theories with fermionic matter content in four dimensions can be inferred from their behavior on $\mathds R^3\times S^1$. In particular, the presence of a mass gap for gauge fluctuations in the decompactification limit can be deduced from the $L$ dependence of the mass gap $m_\sigma$ at finite $L$. If we can assume that the behavior of $m_\sigma$ as a function of $L$ is persistent for all circle sizes, then $m_\sigma$ would only flow down to zero (and hence give us a conformal theory) if and only if it decreases with increasing $L$. Hence, the possibility of calculating $m_\sigma$ in the regime $L\to0$ gives us a criterion for conformality. In short, the theory in the decompactification limit is conformal (confining) if the mass gap is a decreasing (increasing) function of the circle size in the semi-classical window. As we will see, for small number of species $n_f$, $m_\sigma$ is an increasing function of $L$ and as $n_f$ is increased beyond some value $n_f^*$ the behavior flips. This critical number of species is then our estimate for the lower bound of the conformal window. 

\FIGURE{
	\centering
		\includegraphics[width=400pt]{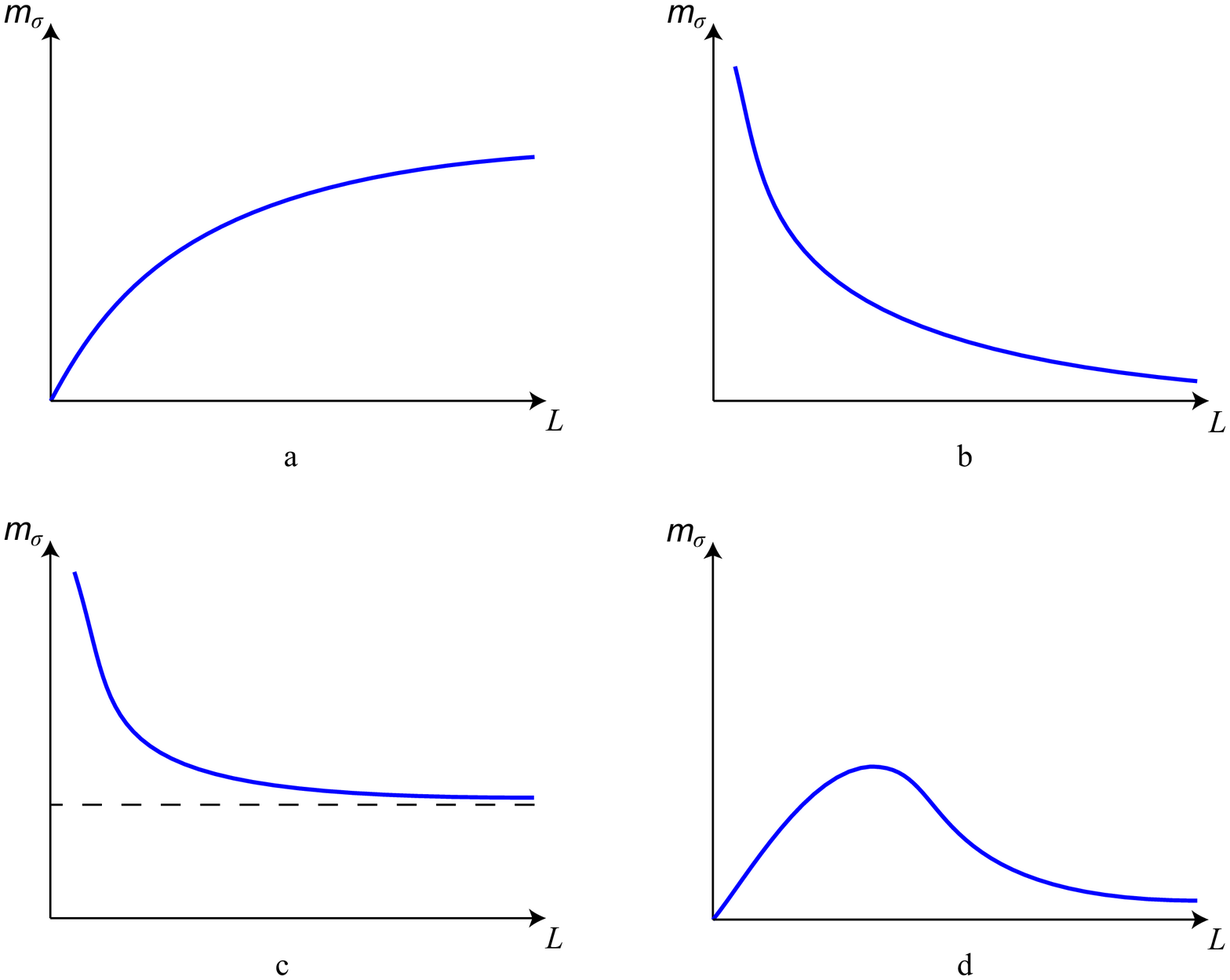}
	\caption{Possible behaviors of the mass gap of gauge fluctuations as a function of the size of the compactified direction. a) Mass gap increasing monotonically and approaching the $R^4$ value asymptotically. b) Mass gap is a monotonically decreasing function of $L$ approaching zero in the decompactification limit. c) Mass gap is a monotonically decreasing function of $L$ but approaches a non-zero asymptote in the decompactification limit. d) Mass gap is an increasing function of $L$ in the semi-classical regime but the behavior flips and approaches zero at large $L$.}
	\label{fig:Classes}
}

However, since the above assumption does not always hold, our prediction will be either an upper or a lower bound on the lower limit of the conformal window. The possibilities are classified in figure \ref{fig:Classes}. If the mass gap of a theory behaves as in case C for some values of $n_f$, the mass gap is a decreasing function of $L$ in the semi-classical regime, therefore, according to the conjecture above, the theory in the decompactification limit should be conformal. Of course this is incorrect because the mass gap asymptotes to a nonzero value. Hence, in this case, the lower boundary of the conformal window is underestimated. A similar argument shows that, in the case D, our procedure is an overestimate.  For more details on peculiarities in the behavior of the mass gap, we refer the reader to \cite{Poppitz:Conf}.

Even though in general our estimates are either an overestimate or an underestimate, in some cases it is possible to predict which class the behavior of the mass gap belongs to, using certain heuristic arguments and hence avoid this discrepancy. We will see examples of this for the case of the fundamental representations.

\subsection{Conformal windows for $SP(2N)$}\label{sec:C_N}

The classic Lie group $C_N$ or $SP(2N)$ has rank $N$, with $N-1$ short roots and one long root. The extended Dynkin diagram is given in \ref{fig:C_N}. 

\FIGURE[ht]{
	\centering
		\includegraphics{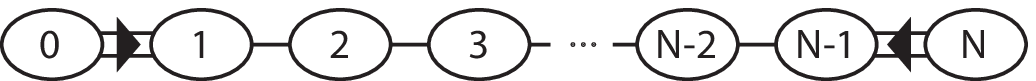}
	\caption{Extended Dynkin diagram of $C_N$. The extended root $\tilde \alpha=\alpha_0$ is proportional to $\lambda_1$.}
	\label{fig:C_N}
}
The vacuum is given by maximally separating the eigenvalues of the holonomy. We note that for the fundamental we would need to add a slight deformation to the effective potential  in order to achieve this structure (as in \cite{Shifman:08}) . The same holds for the 2-index anti-symmetric representation when $n_f=1$. For theories with $1 \leq n_f < 5.5$ symmetric fermions, and $2 \leq n_f < 5.5$ anti-symmetric fermions, center stabilizing double-trace deformations are not needed. In these cases, fermions endowed with periodic boundary conditions  are capable of stabilizing  center-symmetric vacuum. \footnote{The situation is flipped for SO theories, that is one does not need deformations for $1 \leq n_f < 5.5$ anti-symmetric and 
$2 \leq n_f < 5.5$  symmetric representation fermions}. Hence, we have:
\begin{equation}
		A_4|_\infty=\text{diag}\left(\frac{(2N-1)\pi}{2N},\frac{(2N-3)\pi}{2N},\cdots,\frac{\pi}{2N},
				-\frac{\pi}{2N},-\cdots\frac{(2N-3)\pi}{2N},-\frac{(2N-1)\pi}{2N}\right).
		\label{eq:C_N-vacuum}
\end{equation}
Since the monopole solutions corresponding to the simple roots are (anti-)self-dual, their action is proportional to $Q$, their topological charge, via:
\begin{equation}
		S_i=\frac{8\pi^2}{g^2} \left|Q_i\right|,
		\label{eq:action-top-charge}
\end{equation}
We also know that a composite object built up of monopoles with multiplicity $\kappa_\alpha$ is equivalent to the $4D$ instanton (see discussion under equation \eqref{eq:Index-final-regulated}). From this we deduce that the action of the monopole associated to the extended root $\tilde \alpha$ is:
\begin{equation}
		\tilde S= \frac{8\pi^2}{g^2} - \sum_{\alpha\in\Delta} \kappa_\alpha S_\alpha,
		\label{eq:affine-action}
\end{equation}
where the first term is the action of the instanton. This gives $S_{\text{short}}=\dfrac{8\pi^2}{g^2N}$ for the action of the fundamental monopoles corresponding to the short simple roots, $\alpha_i\;1\leq i\leq N-1$, and $S_{\text{long}}=\dfrac{4\pi^2}{g^2N}$ for the long simple root $\alpha_N$ and $\tilde \alpha$. Now, if we had no matter content in our theory, the monopoles would be a source of mass for the dual photons, however, in the presence of fermions, the monopole operators get insertions of fermion fields attached to them:
 \begin{align}
				M_j&=e^{-S_\text{short}}e^{2i\alpha_j\cdot\sigma}\psi^{I_j} &   1\leq &j \leq N-1     &    \text{short simple roots},&\notag\\
				M_j&=e^{-S_\text{long}}e^{i\alpha_j\cdot\sigma}\psi^{I_j}   &          j&=0,N         &    \text{long simple root and } &\tilde\alpha,
		\label{eq:C_N-Mon-Ops}
\end{align}
where $I_j$ is the index of the monopole associated to $\alpha_j$, calculated in the $L\to 0$ limit from equation \eqref{eq:Index-small-L}. Therefore, unless $I_j$ is zero, the monopole operator $M_j$ cannot generate mass for the dual photon combination $\alpha_j\cdot \sigma$, and we must look to higher order operators. We will further elaborate on this point in the examples below.

\subsubsection{$SP(2N)$ with adjoint fermions}

The case of the adjoint representation is similar among all  simple Lie groups, with every monopole inheriting $2$ zero modes from the instanton:
\begin{equation}
		[I_0,I_1,\ldots,I_N]=[2,2,\ldots,2]
		\label{eq:C_N-adj-ind}
\end{equation} 
The mass gap is hence generated, to leading order, through composite objects called bions, which are comprised of one monopole $M_j$ and an adjacent anti-monopole $M_{j\pm1}$ \cite{Unsal:07}. The bions can be schematically depicted as links between the roots in the Dynkin diagram, figure \ref{fig:C_N}. There are $N$ such links, which would give mass to all $N$ dual gauge fields:
\begin{equation}
				B_j=M_j\overline{M_{j+1}}=\left\{
				\begin{aligned}
							e&^{-2S_\text{short}}e^{2i(\alpha_j-\alpha_j+1)\cdot\sigma} 									 &          1&\leq j \leq N-2,     \\
							e&^{-(S_\text{short}+S_\text{long})}e^{i(\tilde\alpha-2\alpha_1)\cdot\sigma} \qquad  &          j&=0,\\	
							e&^{-(S_\text{short}+S_\text{long})}e^{i(2\alpha_{N-1}-\alpha_N)\cdot\sigma}   &          j&=N-1.
				\end{aligned}\right.	
				\label{eq:C_N-adj-ops}
\end{equation}
This gives us the mass of the dual photons in terms of the coupling constant $g^2$. In order to see how this mass depends on the circle size $L$, we define the strong scale $\Lambda$, by using the one-loop beta function:
\begin{equation}
		(\Lambda L)^{\beta_0}=e^{\frac{-8\pi^2}{g^2}}=e^{- N S_{\text{short}}}=e^{-2N S_{\text{long}}},
		\label{eq:C_N-Strong-Scale}
\end{equation}
where $\beta_0=\frac{11}3 C_2(G)-\frac23 T(R)n_f$. 
From this, one can read off the mass of the dual photons. Here, however, we note that at large $N$ most of the fields become massive at order $\exp -2S_\text{short}$. Therefore, in this limit the generated mass gap for $N-2$ of the photons is:
\begin{equation}
		m_\sigma\sim\frac1L e^{-S_{\text{short}}}=\Lambda(\Lambda L)^{\frac{\beta_0}{N}-1}
				=\Lambda (\Lambda L)^{\frac1{3N}\left(8N+11-2n_f(N+1)\right)},
		\label{eq:C_N-adj-MG}
\end{equation} 
where we have used \eqref{eq:C_N-Strong-Scale} for the definition of the strong scale $\Lambda$. From this we obtain our estimate for the conformality bounds:
\begin{equation}
		\frac12\left(8+\frac3{N+1}\right)<n_f<\frac{11}2.
		\label{eq:C_N-adj-bound}
\end{equation}

\subsubsection{$SP(2N)$ with $n_f$ fundamental fermions}\label{sec:C_N-fund}

In this case\footnote{We are only interested in theories with an even number of species in order to avoid the Witten anomaly \cite{Witten:82}. However, we will derive our formulas for general $n_f$ for overall consistency.}, the only monopole with zero modes is the one associated to the long simple root $\alpha_N$, which carries all the $n_f$ zero modes of the instanton:
\begin{equation}
		[I_0,I_1,\ldots,I_{N-1},I_N]=[0,0,\ldots,0,1].
		\label{eq:C_N-fund-ind}
\end{equation}
Therefore, the mass gap is generated through the $N-1$ monopole operators at order $e^{-S_{\text{short}}}$ and the monopole operator associated to the root $\tilde\alpha$ at order $e^{-S_{\text{long}}}$, given as:
\begin{equation}
		\begin{split}
				M_j&=e^{-\frac{8\pi^2}{g^2N}}e^{2i\alpha_j\cdot\sigma} \qquad   1\leq j \leq N-1, \\
				M_0&=e^{-\frac{4\pi^2}{g^2N}}e^{i\tilde\alpha\cdot\sigma}.
		\end{split}
		\label{eq:C_N-fund-ops}
\end{equation}
With the aid of equation \eqref{eq:C_N-Strong-Scale}, we derive the characteristic mass gap:
\begin{align}
		m_{\alpha_j\cdot\sigma}&\sim\frac1L e^{-\frac{S_{\text{short}}}{2}}=\Lambda(\Lambda L)^{\frac{\beta_0}{2N}-1}
				=\Lambda (\Lambda L)^{\frac1{6N}\left(5N+11-n_f\right)}\qquad   1\leq j \leq N-1,  \notag\\
		m_{\tilde\alpha\cdot\sigma}&\sim\frac1L e^{-\frac{S_{\text{long}}}{2}}=\Lambda(\Lambda L)^{\frac{\beta_0}{4N}-1}
				=\Lambda (\Lambda L)^{\frac1{12N}\left(-N+11-n_f\right)}. 		\label{eq:C_N-fund-MG}
\end{align}
For our approximation of the conformal window, we only look at the contribution of the short roots which provide the dominant contribution in the large $N$ limit. We see that for fixed $N$ and small $n_f$ the generated mass grows as we increase L. As we increase $n_f$ past the critical value $5N+11$, the behavior flips and decreases with increasing circle size $L$. Therefore, according to the assertion in section \ref{sec:conf-bound}, our procedure tells us that the conformality bound is:
\begin{equation}
		5N+11<n_f<11(N+1).
		\label{eq:C_N-fund-bound}
\end{equation}

It is fruitful at this point to compare the results of the fundamental representation to the adjoints. From counting arguments we know that perturbatively  one adjoint fermion of $SP(2N)$ behaves like $2N+2$  fundamentals. Therefore, from \eqref{eq:C_N-adj-bound} we would expect the lower bound of the conformal window to be $8N+11$ as opposed to $5N+11$. We believe this discrepancy is caused by our inability to differentiate cases A and C in figure \ref{fig:Classes} and predict that theories that fall in the region of the mismatch ($5N+11<n_f<8N+11$) belong to the anomalous case C. These are theories that are confining even though their mass gap decreases with increasing $L$ in the semiclassical limit. Therefore, our final estimate for the conformal window is: 
\begin{equation}
		8N+11<n_f<11(N+1).
		\label{eq:C_N-fund-bound-final}
\end{equation}

\subsubsection{$SP(2N)$ with 2-index anti-symmetric fermions}\label{sec:C_N-anti}

This representation can be considered as the compliment to the defining representation, in that it contains all the short roots whereas the defining representation contained all the long roots (with their length divided by $2$). The index of the monopoles is given as:
\begin{equation}
		[I_0,I_1,I_2,\ldots,I_{N-1},I_N]=[0,2,2,\ldots,2,0].
		\label{eq:C_N-anti-ind}
\end{equation}
We see that the distribution of the zero modes is also the compliment of the defining representation where only the long root had a nonzero index. The mass gap is now generated through a combination of monopoles and bions:
\begin{align}
		M_j&=e^{-S_\text{long}}e^{i\alpha_j\cdot\sigma} 																			 &			 j&=0,N 				&&\text{long roots}, \notag\\
		B_j&=M_j\overline{M_{j+1}}=e^{-2S_\text{short}}e^{2i(\alpha_j-\alpha_j+1)\cdot\sigma}  &   1\leq &j \leq N-2 \qquad  &&\text{links } \text{between short roots}.
		\label{eq:C_N-anti-ops}
\end{align}
In the large $N$ limit, the predominant effect is that of the bions. Hence, the mass gap for all but 2 of the photons is:
\begin{equation}
		m_\sigma\sim\frac1L e^{-S_{\text{short}}}=\Lambda(\Lambda L)^{\frac{\beta_0}{N}-1}
				=\Lambda (\Lambda L)^{\frac1{3N}\left(8N+11-2n_f(N-1)\right)},
		\label{eq:C_N-anti-MG}
\end{equation}
where again we have used \eqref{eq:C_N-Strong-Scale} to define $\Lambda$. The estimate for the conformality window is then:
\begin{equation}
		\frac12\left(8+\frac{19}{N-1}\right)<n_f<\frac{11}2\frac{N+1}{N-1}.
		\label{eq:C_N-anti-bound}
\end{equation}

\subsection{Conformal windows for $SO(2N+1)$}\label{sec:B_N}

The structure of the Lie group $B_N$ or $SO(2N+1)$ is similar to that of $C_N$, as they share the same Coxeter graph. However, the different lengths of the roots makes the extended Dynkin diagram look very different. As can be seen from figure \ref{fig:B_N}, the rank of the group is $N$, with 1 short root $\alpha_N$ and $N-1$ long roots. The extended root $\tilde\alpha$ is in this case a long root proportional to the fundamental weight $\lambda_2$.
\FIGURE{
	\centering
		\includegraphics{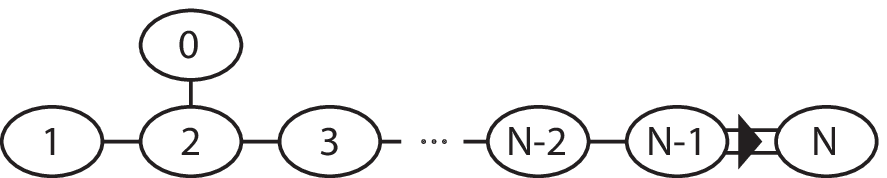}
	\caption{Extended Dynkin diagram of $B_N$. The root $\tilde \alpha=\alpha_0$ is proportional to $\lambda_2$.}
	\label{fig:B_N}
}

The structure of the vacuum is identical to the $SU(2N+1)$ case and is given by:
\begin{equation}
		A_4|_\infty=\text{diag}\left(\frac{2N\pi}{2N+1},\frac{2(N-1)\pi}{2N+1},\cdots,\frac{2\pi}{2N+1}
				,\,0\:,-\frac{2\pi}{2N+1},\cdots,\frac{-2(N-1)\pi}{2N+1},\frac{-2N\pi}{2N+1}\right).
		\label{eq:B_N-vacuum}
\end{equation}
Similar to the $SP(2N)$ case, a slight deformation of the effective potential is need to attain this structure in the case of the fundamental representation as well as the 2-index symmetric representation with $n_f=1$. The same holds for $SO(2N)$. From this, we calculate the action of the monopoles:
\begin{align}
		S_\text{long}&=\frac{8\pi^2}{g^2(2N+1)}&\text{long roots},&\notag\\
		S_\text{short}&=S_{\tilde\alpha}=\frac{16\pi^2}{g^2(2N+1)}&\text{short roots }&\text{and }\tilde\alpha.
		\label{eq:B_N-monopole-action}
\end{align}
We will also use the strong scale derived from the one loop beta function:
\begin{equation}
		(\Lambda L)^{\beta_0}=e^{\frac{-8\pi^2}{g^2}}=e^{- \frac12 (2N+1) S_{\text{short}}}=e^{-(2N+1) S_{\text{long}}}.
		\label{eq:B_N-Strong-Scale}
\end{equation}

\subsubsection{$SO(2N+1)$ with adjoint fermions}

Similar to the adjoint representation of all other gauge groups, all the monopoles carry an equal number of zero modes. We have:
\begin{equation}
		[I_0,I_1,\ldots,I_N]=[2,2,\ldots,2].
		\label{eq:B_N-adj-ind}
\end{equation}
Hence, the leading operators responsible for the mass gap are the bions. Again, the best way to visualize these operators is to think of them as the links between the roots of the extended Dynkin diagram (figure \ref{fig:B_N}).
\begin{equation}
				B_j=\left\{
				\begin{aligned}
						&M_j\overline{M_{j+1}}=e^{-2S_\text{long}}e^{i(\alpha_j-\alpha_j+1)\cdot\sigma} 					   	  &          1&\leq j \leq N-2,     \\
						&M_{N-1}\overline{M_N}=e^{-(S_\text{short}+S_\text{long})}e^{i(\alpha_{N-1}-2\alpha_N)\cdot\sigma} \qquad  &          j&=N-1,\\	
						&M_0\overline{M_2}=e^{-(S_\text{short}+S_\text{long})}e^{i(\tilde\alpha-\alpha_2)\cdot\sigma}     			   &          j&=0.
				\end{aligned}\right.	
				\label{eq:B_N-adj-ops}
\end{equation}
Noting that the $N-2$ of the photons receive mass at scale $\exp -S_\text{long}$, the estimate of the mass gap for large $N$ is:
\begin{equation}
		m_\sigma\sim\frac1L e^{-S_{\text{long}}}=\Lambda(\Lambda L)^{\frac{\beta_0}{2N+1}-1}
				=\Lambda (\Lambda L)^{\frac1{3N}\left(16N-14-2n_f(2N-1)\right)},
		\label{eq:B_N-adj-MG}
\end{equation} 
where we have used equation \ref{eq:B_N-Strong-Scale}. Our estimate for the conformality window is then:
\begin{equation}
		4-\frac3{2N-1}<n_f<\frac{11}2.
		\label{eq:B_N-adj-bound}
\end{equation}

\subsubsection{$SO(2N+1)$ with fundamental fermions}

The fundamental representation of $SO(2N+1)$ is very similar to the $2$-index antisymmetric representation of $SP(2N)$ discussed in section \ref{sec:C_N-anti}. Here, also, the representation is comprised of all the short roots. The index of the monopoles is hence:
\begin{equation}
		[I_0,I_1,\ldots,I_{N-1},I_N]=[0,0,\ldots,0,2],
		\label{eq:B_N-fund-ind}
\end{equation}
which implies that the monopoles associated to the long roots are responsible for generating a mass gap. Their explicit form is given as:
\begin{equation}
		M_j=\left\{
		\begin{aligned}
				&e^{-\frac{8\pi^2}{g^2(2N+1)}}e^{i\alpha_j\cdot\sigma} \qquad   1\leq j \leq N-1, \\
				&e^{-\frac{16\pi^2}{g^2(2N+1)}}e^{i\tilde\alpha\cdot\sigma}\qquad\; j=0,
		\end{aligned}\right.
		\label{eq:B_N-fund-ops}
\end{equation}
which are mostly at the scale $\exp -S_\text{long}$. The generated mass gap is then:
\begin{equation}
		m_{\alpha_j\cdot\sigma}\sim\left\{
		\begin{aligned}
				&\frac1L e^{-\frac12 S_\text{long}}=\Lambda(\Lambda L)^{\frac{\beta_0}{2(2N+1)}-1}=\Lambda(\Lambda L)^{\frac1{6(2N+1)} (10N-17-2n_f)} \qquad   1\leq j \leq N-1,\\
				&\frac1L e^{-\frac12 S_{\tilde\alpha}}=\Lambda(\Lambda L)^{\frac{\beta_0}{(2N+1)}-1}=\Lambda(\Lambda L)^{\frac1{3(2N+1)} (16N-14-2n_f)}\qquad\quad j=0,
		\end{aligned}
		\right.
		\label{eq:B_N-fund-MG}
\end{equation}
where we have made use of the strong scale defined in equation \eqref{eq:B_N-fund-MG}. Taking only the predominant effect of the long roots into consideration, we derive our estimate of the conformality window:
\begin{equation}
		\frac12 (10N-17)<n_f<\frac{11}2(2N-1).
		\label{eq:B_N-fund-bound}
\end{equation}

The discussion of section \ref{sec:C_N-fund} applies equally here. In this case, one adjoint fermion is the equivalent of $2N-1$ fundamentals, therefore, comparing to the case of the adjoints we predict that theories with $5N-8.5<n_f<8N-7$ belong to the anomalous case C and the true conformal window is:

\begin{equation}
		8N-7<n_f<\frac{11}2(2N-1).
		\label{eq:B_N-fund-bound-final}
\end{equation}

\subsubsection{$SO(2N+1)$ with 2-index symmetric fermions} \label{B_N-sym}

This representation of $B_N$ is comprised of all the roots, as well as the weights which have twice the length of the short roots. This means that the index of the monopoles associated to the long roots is the same as the adjoint representation, while the index of the monopoles associated to the short roots and the root $\tilde\alpha$ is 3 times that of the adjoint representation:
\begin{equation}
		[I_0,I_1,\ldots I_{N-1},I_N]=[6,2,\ldots,2,6].
		\label{eq:B_N-sym-ind}
\end{equation}
The leading operators with zero magnetic charge and hence no zero modes are magnetic bions and quartets:
\begin{align}
		B_j&=M_j\overline{M_{j+1}}=e^{-2S_\text{long}}e^{i(\alpha_j-\alpha_j+1)\cdot\sigma} 					   	\qquad\qquad          1\leq j \leq N-2,     \notag\\
		Q_0&=M_0\overline{M_2}^3=e^{-(S_\text{short}+3S_\text{long})}e^{i(\tilde\alpha-3\alpha_2)\cdot\sigma},\notag\\
		Q_N&=M_{N-1}^3\overline{M_N}=e^{-(S_\text{short}+3S_\text{long})}e^{i(3\alpha_{N-1}-2\alpha_N)\cdot\sigma}.
		\label{eq:B_N-sym-ops}
\end{align}
Again, these composite operators can be visualized as the links between the roots on the extended Dynkin diagram. The operators $B_j$ connect the $\alpha_j$ to $\alpha_{j+1}$ and the operators $Q_0$ and $Q_N$ connect  $\tilde\alpha$ to $\alpha_2$ and $\alpha_{N-1}$ to $\alpha_N$ respectively. 

This structure is more complex than the previous cases, however, we expect the magnetic bions to have the predominant effect at large N. Therefore, in this limit the generated mass gap would be:
\begin{equation}
		m_\sigma\sim\frac1L e^{-S_{\text{long}}}=\Lambda(\Lambda L)^{\frac{\beta_0}{2N+1}-1}
				=\Lambda (\Lambda L)^{\frac1{3N}\left(16N-14-2n_f(2N+3)\right)}.
		\label{eq:B_N-sym-MG}
\end{equation} 
The predicted range of conformality is then:
\begin{equation}
		4-\frac{19}{2N+3}<n_f<\frac{11}2.
		\label{eq:B_N-sym-bound}
\end{equation}

\subsection{Conformal windows for $SO(2N)$}\label{sec:D_N}

The  main differentiating feature of the Lie algebra $D_N$ or $SO(2N)$ from the other classical Lie algebras is the branching of the Dynkin diagram on $\alpha_{N-2}$ (figure \ref{fig:D_N}). 
\FIGURE{
	\centering
		\includegraphics{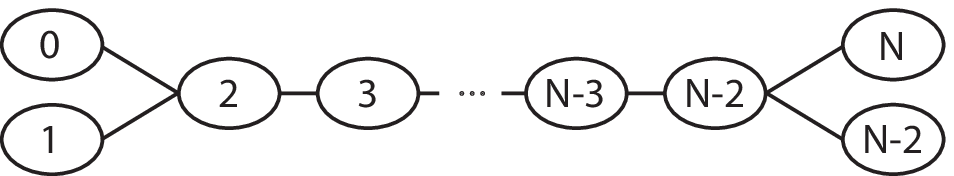}
	\caption{Extended Dynkin diagram of $D_N$. The root $\tilde \alpha=\alpha_0$ is proportional to $\lambda_2$.}
	\label{fig:D_N}
}

This  means that there is, a priori, no specific order on the weights of the fundamental representation. Equivalently, one can say that elements of the Weyl group that fix the Weyl chamber, do not necessarily fix the vacuum  of the theory. Here we make the choice:
\begin{equation}
	w_{N+1}(A_4|_\infty)<0<w_N(A_4|_\infty),
	\label{eq:D_N-choice}
\end{equation} 
which fixes the vacuum structure to:
\begin{equation}
		A_4|_\infty=\text{diag}\left(\frac{(2N-1)\pi}{2N},\frac{(2N-3)\pi}{2N},\cdots,\frac{\pi}{2N}
				,-\frac{\pi}{2N},-\cdots,-\frac{(2N-3)\pi}{2N},-\frac{(2N-1)\pi}{2N}\right),
		\label{eq:D_N-vacuum}
\end{equation}
from which we calculate the action of the monopole operators:
\begin{equation}
		S_i=\left\{
		\begin{aligned}
			&\frac{4\pi^2}{g^2 N} &&1\leq j \leq N-1,\\
			&\frac{8\pi^2}{g^2 N}&&j=0,N.
		\end{aligned}
		\right.
		\label{eq:D_N-monopole-action}
\end{equation}
The strong scale derived from the one loop beta function is:
\begin{equation}
		(\Lambda L)^{\beta_0}=e^{\frac{-8\pi^2}{g^2}}=e^{-N S_0}=e^{- 2N S_j} \qquad 1\leq j \leq N-1.
		\label{eq:D_N-Strong-Scale}
\end{equation}

\subsubsection{$SO(2N)$ with adjoint fermions}

As previously mentioned, the zero mode distribution in the adjoint representation of all gauge groups is always equal across the board:
\begin{equation}
		[I_0,I_1,\ldots,I_N]=[2,2,\ldots,2].
		\label{eq:D_N-adj-ind}
\end{equation}
With magnetic bions generating the mass gap to leading order. In this case the relevant operators are:
\begin{equation}
				B_j=\left\{
				\begin{aligned}
						&M_j\overline{M_{j+1}}=e^{-2S_j}e^{i(\alpha_j-\alpha_j+1)\cdot\sigma} 					   	  &          1&\leq j \leq N-2,     \\
						&M_{N-2}\overline{M_N}=e^{-(S_{N-2}+S_N)}e^{i(\alpha_{N-2}-\alpha_N)\cdot\sigma} \qquad  &          j&=N,\\	
						&M_0\overline{M_2}=e^{-(S_0+S_2)}e^{i(\tilde\alpha-\alpha_2)\cdot\sigma}     			   &          j&=0.
				\end{aligned}\right.	
				\label{eq:D_N-adj-ops}
\end{equation}
Hence, our estimate of the mass gap for large $N$ is:
\begin{equation}
		m_\sigma\sim\Lambda(\Lambda L)^{\frac{\beta_0}{2N}-1}
				=\Lambda (\Lambda L)^{\frac1{3N}\left(16N-22-2n_f(2N-2)\right)}.
		\label{eq:D_N-adj-MG}
\end{equation} 
Where we have used equation \ref{eq:D_N-Strong-Scale}. The conformality bound is given by:
\begin{equation}
		4-\frac3{2N-2}<n_f<\frac{11}2.
		\label{eq:D_N-adj-bound}
\end{equation}

\subsubsection{$SO(2N)$ with fundamental fermions}

Similar to the other classical Lie algebras, the zero modes of the fundamental representation are localized to a single monopole. With equation \eqref{eq:D_N-choice} determining our base $\Delta$, the only nonzero index is $I_N$:
\begin{equation}
		[I_0,I_1,\ldots,I_{N-1},I_N]=[0,0,\ldots,0,2].
		\label{eq:D_N-fund-ind}
\end{equation}
Hence, the monopole operators are responsible for generating the mass gap. The mass of the dual photons is estimated at:
\begin{equation}
	m_{\alpha_j\cdot\sigma}\sim\frac1L e^{-\frac12 S_j}=\left\{
	\begin{aligned}
		&\Lambda(\Lambda L)^{\frac{\beta_0}{4N}-1}=\Lambda(\Lambda L)^{\frac13(10N-22-2n_f)} &1\leq& j \leq N-1,\\
		&\Lambda(\Lambda L)^{\frac{\beta_0}{2N}-1}=\Lambda(\Lambda L)^{\frac13(16N-22-2n_f)} && j=0.\\
	\end{aligned}
	\right.
	\label{eq:D_N-fund-MG}
\end{equation}
As before, we only consider the predominant effect in the calculation of the conformality window:
\begin{equation}
		5N-11<n_f<\frac{11}2(2N-2).
		\label{eq:D_N-fund-bound}
\end{equation}

Just as in the fundamental fermions of $SP(2N)$ and $SO(2N+1)$, we find a discrepancy between this result and that of the adjoints. Substituting $2N-2$ fundamental fermions for each adjoint in \eqref{eq:D_N-adj-bound}, we see a mismatch in the conformal window in the range $5N-11<n_f<8N-11$. As before, we predict that these theories belong to the anomalous case C and the true conformal window is:
\begin{equation}
		8N-11<n_f<\frac{11}2(2N-2).
		\label{eq:D_N-fund-bound-final}
\end{equation}

\subsubsection{$SO(2N)$ with 2-index symmetric fermions}\label{sec:D_N-sym}

With our choice of the base $\Delta$ (equation \eqref{eq:D_N-choice}), this case is almost identical to section \ref{B_N-sym}. The representation is comprised of the roots plus twice the weights of the fundamental representation. The distribution of the zero modes is then:
\begin{equation}
		[I_0,I_1,\ldots I_{N-1},I_N]=[6,2,\ldots,2,6].
		\label{eq:D_N-sym-ind}
\end{equation}
The analysis is identical to section \ref{B_N-sym} with magnetic bions and quartets generating the mass gap:
\begin{align}
		B_j&=M_j\overline{M_{j+1}}=e^{-2S_j}e^{i(\alpha_j-\alpha_j+1)\cdot\sigma} 					   	\qquad\qquad          1\leq j \leq N-2,     \notag\\
		Q_0&=M_0\overline{M_2}^3=e^{-(S_0+3S_2)}e^{i(\tilde\alpha-3\alpha_2)\cdot\sigma},\notag\\
		Q_N&=M_{N-1}^3\overline{M_N}=e^{-(S_N+3S_{N-1})}e^{i(3\alpha_{N-1}-2\alpha_N)\cdot\sigma}.
		\label{eq:D_N-sym-ops}
\end{align}
Again, these composite operators can be visualized as the links between the roots on the extended Dynkin diagram (figure \ref{fig:D_N}). In the large $N$ limit, we expect the magnetic bions to have the predominant effect, therefore, in this limit the generated mass gap would be:
\begin{equation}
		m_\sigma\sim\Lambda(\Lambda L)^{\frac{\beta_0}{2N}-1}
				=\Lambda (\Lambda L)^{\frac1{3N}\left(16N-22-2n_f(2N+2)\right)}.
		\label{eq:D_N-sym-MG}
\end{equation} 
The predicted range of conformality is:
\begin{equation}
		4-\frac{19}{2N+2}<n_f<\frac{11}2.
		\label{eq:D_N-sym-bound}
\end{equation}

\section{Comparison with other estimates of the conformal window}\label{sec:Comparisons}

The results of section \ref{sec:conf-bound} are given in tables \ref{tab:C_N-fund} to \ref{tab:SO(N)-sym}, for several values of $N$ (referred to as Deformation Theory). Since the  results for $SO(2N)$ and $SO(2N+1)$ are similar, they have been combined and appear in the table under $SO(N)$. For comparison we have also included the results from three other approaches:  the ladder approximation using truncated Schwinger-Dyson equations, the NSVZ-inspired approach with both $\gamma=1$ and $\gamma=2$ results (for a review of these see \cite{Pagels:75,Appelquist:1988yc,Dietrich:2006cm,Sannino:08,Sannino:09}), as well as the estimates from the worldline formalism\cite{Armoni:09}. We also not that, for the cases at hand, there are no lattice results available that we are aware of.

We note that the comparison here bears a striking resemblance to a similar comparison carried out in \cite{Poppitz:Conf} for the gauge group $SU(N)$. For all gauge groups the following holds. For the two indexed representations, the deformation theory approximation is close to the Schwinger-Dyson equation and the worldline formalism and is relatively far from the NSVZ-inspired estimates.

 We also note that if we were to take the bare result of our calculations \eqref{eq:C_N-fund-bound},\eqref{eq:B_N-fund-bound} and \eqref{eq:D_N-fund-bound} without accounting for the discrepancy coming from the switch from case A to C, then our approximation would have been much closer to the NSVZ-inspired estimate with $\gamma=2$  and would have been far from the $\gamma=1$ estimate and farther still from the Schwinger-Dyson and worldline formalism numbers. However, with this correction, our results are again close to the Schwinger-Dyson equation and the worldline formalism and are far from the NSVZ-inspired estimates. Tables \ref{tab:C_N-fund} and \ref{tab:SO(N)-fund} demonstrate our results both before and after this correction, respectively under columns A and A+C.


\TABLE{
\begin{tabular}{|c||c|c|c|c|c|c|}
		\hline
\multirow{2}{*}{N}	&	Deformation			&		Ladder (SD)						&	\multicolumn{2}{|c|}{	NSVZ-inspired}	&	Worldline	&		Asymptotic\\\cline{4-5}
							&		Theory				&	approx.					&		 $\gamma=1$ 								& $\gamma=2$					&		Formalism							& 	Freedom \\\hline
		2					&		4.5						&		4.15					&		3.67												&		2.75							&4												&		5.5\\
		3					&		4.37					&		4.15					&		3.67												&		2.75							&4												&		5.5 \\
		4					&		4.3						&		4.15					&		3.67												&		2.75							&4												&		5.5 \\
		5					&		4.25					&		4.15					&		3.67												&		2.75							&4												&		5.5 \\
		10				&		4.13					&		4.15					&		3.67												&		2.75							&4												&		5.5 \\
		$\infty$	&		4							&		4.15					&		3.67												&		$2.75$						&4												&		5.5 \\
		\hline
\end{tabular}
\caption{Lower boundary estimate of the conformal window for $SP(2N)$ with adjoint fermions.}
\label{tab:C_N-adj}
}

\TABLE{
\begin{tabular}{|c||c|c|c|c|c|c|c|}
		\hline
\multirow{2}{*}{N}	&	\multicolumn{2}{|c|}{Def. Theory}	&		Ladder 	&	\multicolumn{2}{|c|}{	NSVZ-inspired}	&	Worldline	&		Asymptotic\\\cline{2-3}\cline{5-6}
							&		A	&	A+C				&	approx.					&		 $\gamma=1$ 								& $\gamma=2$					&		Formalism							& 	Freedom \\\hline
		2					&		21&	27						&		23.7									&		22													&		16.5	  	&24															&		33	\\
		3					&		26&35						&		31.7									&		29.3												&		22		  	&32															&		44 \\
		4					&		31&43						&		39.7									&		36.6												&		27.5			&40															&		55 \\
		5					&		36&51						&		47.7									&		44													&		33				&48															&		66 \\
		10				&		61&91						&		87.7									&		80.6												&		60.5			&88															&		121 \\
		$\infty$	&		5N+11	&	8N+11					&		8N+7.7								&		$\frac{22}3(N+1)$						&		$\frac{11}2(N+1)$		&8N+8									&		11(N+1) \\
		\hline
\end{tabular}
\caption{Lower boundary estimate  of the conformal window for $SP(2N)$ with fundamental fermions.}
\label{tab:C_N-fund}
}

\TABLE{
\begin{tabular}{|c||c|c|c|c|c|c|}
		\hline
\multirow{2}{*}{N}	&	Deformation			&		Ladder (SD)						&	\multicolumn{2}{|c|}{	NSVZ-inspired}	&	Worldline	&		Asymptotic\\\cline{4-5}
							&		Theory			&	approx.					&		 $\gamma=1$ 								& $\gamma=2$					&		Formalism							& 	Freedom \\\hline
		2					&		13.5				&		12.2					&		11													&		8.25							&12												&		16.5\\
		3					&		8.75				&		8.18					&		7.33												&		5.5								&8												&		11 \\
		4					&		7.16				&		7.17					&		6.11												&		4.58							&6.67											&		9.17 \\
		5					&		6.375				&		6.37					&		5.5													&		4.12							&6												&		8.25 \\
		10				&		5.05				&		5.05					&		4.48												&		3.36							&4.89											&		6.72 \\
		$\infty$	&		4						&		4.15					&		3.67												&		2.75							&4												&		5.5 \\
		\hline
\end{tabular}
\caption{Estimate of the lower boundary of the conformal window for $SP(2N)$ with 2-index \mbox{anti-symmetric} fermions.}
\label{tab:C_N-anti}
}


\TABLE{
\begin{tabular}{|c||c|c|c|c|c|c|}
		\hline
\multirow{2}{*}{N}	&	Deformation			&		Ladder (SD)						&	\multicolumn{2}{|c|}{	NSVZ-inspired}	&	Worldline	&		Asymptotic\\\cline{4-5}
							&		Theory				&	approx.					&		 $\gamma=1$ 								& $\gamma=2$					&		Formalism							& 	Freedom \\\hline
		6						&		3.25				&		4.15									&		3.67												&		2.75			&4													&		5.5 \\
		7						&		3.4					&		4.15									&		3.67												&		2.75			&4													&		5.5 \\
		8						&		3.5					&		4.15									&		3.67												&		2.75			&4													&		5.5 \\
		9						&		3.57				&		4.15									&		3.67												&		2.75			&4													&		5.5 \\
		10					&		3.62				&		4.15									&		3.67												&		2.75			&4													&		5.5	\\
		$\infty$		&		4						&		4.15									&		3.67												&		2.75			&4													&		5.5 \\
		\hline
\end{tabular}
\caption{Lower boundary estimate of the conformal window for $SO(N)$ with adjoint fermions.}
\label{tab:SO(N)-adj}
}

\TABLE{
\begin{tabular}{|c||c|c|c|c|c|c|c|}
		\hline
\multirow{2}{*}{N}	&	\multicolumn{2}{|c|}{Def. Theory}	&		Ladder 	&	\multicolumn{2}{|c|}{	NSVZ-inspired}	&	Worldline	&		Asymptotic\\\cline{5-6}\cline{2-3}
							&		A&A+C				&	approx.					&		 $\gamma=1$ 								& $\gamma=2$					&		Formalism				& 	Freedom \\\hline
		6					&		4&13							&		16.2					&		14.7												&		11								&16									&		22 \\
		7					&		6.5&17						&		20.2					&		18.3												&		13.75							&20									&		27.5 \\
		8					&		9&21							&		24.2					&		22													&		16.5							&24									&		33 \\
		9					&		11.5&25					&		28.2					&		25.7												&		19.25							&28									&		38.5 \\
		10				&		14&29						&		32.2					&		29.3												&		22								&32									&		44	\\
		$\infty$	&		$\frac52 N-11$&$4N-11$	&		4N-7.76				&		$\frac{11}3(N-2)$						&		$\frac{11}4(N-2)$	&4N-8								&		$\frac{11}2(N-2)$ \\
		\hline
\end{tabular}
\caption{Lower boundary estimate of the conformal window for $SO(N)$ with fundamental fermions.}
\label{tab:SO(N)-fund}
}

\TABLE{
\begin{tabular}{|c||c|c|c|c|c|c|}
		\hline
\multirow{2}{*}{N}	&	Deformation			&		Ladder (SD)						&	\multicolumn{2}{|c|}{	NSVZ-inspired}	&	Worldline	&		Asymptotic\\\cline{4-5}
							&		Theory				&	approx.					&		 $\gamma=1$ 								& $\gamma=2$					&		Formalism							& 	Freedom \\\hline
		6						&		1.6					&		2.1										&		1.8													&		1.4				&2												&		2.75 \\
		7						&		1.9					&		2.3										&		2.0													&		1.5				&2.22											&		3.05 \\
		8						&		2.1					&		2.5										&		2.2													&		1.6				&2.4											&		3.3 \\
		9						&		2.3					&		2.7										&		2.3													&		1.7				&2.55											&		3.5 \\
		10					&		2.4					&		2.8										&		2.4													&		1.8				&2.67											&		3.67	\\
		$\infty$		&		4						&		4.15									&		3.67												&		2.75			&4												&		5.5 \\	\hline
\end{tabular}
\caption{Estimate of the lower boundary of the conformal window for $SO(N)$ with 2-index symmetric fermions.}
\label{tab:SO(N)-sym}
}

\section{Conclusion}\label{sec:Conclusion}

A quick look at the tables in section \ref{sec:Comparisons} as well as a comparison with the $SU(N)$ results from \cite{Poppitz:Conf} shows that there are no surprises in our results: across the board the conformal bounds for all classical gauge groups are more or less the same, especially in the asymptotic limit. Of course, this was to be expected since for large $N$, there exists a  nonperturbative orientifold equivalence which relates 
neutral sectors of  $SO$, $Sp$ and $SU$ theories \cite{Armoni:2003gp,Unsal:2006pj}. In particular, this equivalence implies the equality for  the onset of the conformal window in the large N limit. Therefore, any hope of distinguishing the different cases rests in the finite $N$ regime. However, the difficulty of deriving the coefficients of monopole and composite monopole operators makes this calculation inaccessible. The same goes for the non-classical Lie groups.

Even though systematic improvement of our results is out of reach, as we showed in the case of the fundamental fermions, it is still possible to predict which class (figure \ref{fig:Classes}) the behavior of the mass gap of any given theory falls under. We hope that in the future, similar arguments  might be able to classify all of the cases considered here and put our predictions in context.

\acknowledgments
We would like to thank E. Poppitz for continuous support throughout the project. We are also greatly indebted to M. \"Unsal for a generous amount of helpful comments and discussions. We also thank A. Armoni for pointing out to us the implications of orientifold equivalence with regard to the conformal window.

\begin{center}
    {\bf APPENDIX}
\end{center}
\appendix
\section{Lie Algebra Preliminaries}\label{app:Lie_Prel}

We follow the notation of \cite{Humphreys}. Let $G$ be a simple gauge group of rank $l$ with $\mathfrak g$ its Lie algebra. Denote $\mathfrak h$ as the Cartan subalgebra, $\Phi$ as the set of all roots and $w^R_i$ as the weights of a representation $R$. 
 The weights correspond to maps $w^R_i:\mathfrak h \to \mathds R$ and are hence elements of $\mathfrak h^*$. Weight spaces are defined as:
\begin{equation}
L_{w_i^R}=\left\{x\in R(\mathfrak g)| \;\forall h \in \mathfrak h,\, h(x)=w_i^R(h)\,x\right\},
\label{eq:weight_spaces}
\end{equation}
and the roots are the weights of the adjoint representation. With this definition it is clear that by $w(h)$ we mean the eigenvalue of h corresponding to the weight $w$. For example if $h=\text{diag}(h_1,\ldots,h_n)$ in the fundamental representation and $w_i$ is the $i$'th weight of the fundamental representation, we have $w_i(h)=h_i$.

We choose a basis for $\Phi$ such that all roots can be written as a sum over the basis elements with coefficients of the same sign. The elements of this basis are called simple roots and form the set $\Delta$ which is called a base. 

Throughout this paper, we will use two properties of Lie algebras. 
\begin{enumerate}
\item The existence of a non-degenerate bilinear form $b:\mathfrak h \times \mathfrak h \to \mathds R$, $b(h_1,h_2)=\text {tr}(h_1\cdot h_2)$ allows us to identify $\mathfrak h$ with $\mathfrak h^*$: to $w\in\mathfrak h^*$ corresponds the unique element $t_w\in\mathfrak h$ satisfying: 
\begin{equation}
w(h)=b(t_w,h) \qquad \forall h\in\mathfrak h.
\label{eq:weight-2-h}
\end{equation}

The bilinear form $b$ also defines\footnote{This is the only place our notation differs from \cite{Humphreys} where the inner product is defined using the Killing form $\kappa$. The relation between the two definitions is given by\[\kappa(t_1,t_2)=\frac{T(G)}{T(F)}b(t_1,t_2).\]}  an inner product on $\mathfrak h^*$ given by: 
\begin{equation}
w_1\cdot w_2=b(t_{w_1},t_{w_2})=w_2(t_{w_1})=w_1(t_{w_2}).
\label{eq:weights_inner_product}
\end{equation}

Using this inner product, we define the coroots as: 
\begin{equation}
\alpha^*=\frac{2\alpha}{|\alpha^2|},
\label{eq:coroots}
\end{equation}
and the fundamental weights $\left\{\lambda_i\right\}$ as the members of the basis dual to the coroots:
\begin{equation}
		\lambda_i\cdot \alpha_j^*=\delta_{ij}.
		\label{eq:fund-weights}
\end{equation}
\item If $\alpha \in \Phi$ and $x_\alpha\in L_\alpha,\;x_\alpha\neq0$, then there exists $y_\alpha\in L_{-\alpha}$ such that $x_\alpha, \;y_\alpha,\; h_\alpha=[x_\alpha,\;y_\alpha]$ span a three dimensional simple subalgebra of L isomorphic to $\mathfrak{sl}(2,\mathds C)$ via: 
\begin{equation}
x_\alpha\to\begin{pmatrix}0&1\\0&0\end{pmatrix},\quad
y_\alpha\to\begin{pmatrix}0&0\\1&0\end{pmatrix},\quad
h_\alpha\to\begin{pmatrix}1&0\\0&-1\end{pmatrix}.
\label{eq:sl(2)_embed}
\end{equation}
Here, the elements $x_\alpha$ and $y_\alpha$ are respectively the raising and lowering operators in the direction of the root $\alpha$. Furthermore, we have:
\begin{equation}
	h_\alpha=\frac{2t_\alpha}{b(t_\alpha,t_\alpha)}=\frac{2}{|\alpha|^2}t_\alpha=t_{\alpha^*}.
	\label{eq:t_to_h}
\end{equation}
\end{enumerate}

In any root system, there is a unique root $\tilde\alpha$ such that: 
\begin{equation}
	y_\sigma(\tilde\alpha)=0 \qquad \forall \sigma\in\Delta,
	\label{eq:alpha-tilde}
\end{equation}
where $y_\sigma$ is the lowering operator defined in property 2 above. The root $\tilde\alpha$ is called the lowest root (for obvious reasons). We define the extended base $\tilde\Delta$ as:
\begin{equation}
	\tilde\Delta=\Delta\cup\left\{\tilde\alpha\right\}.
	\label{eq:extended-base}
\end{equation}

It is conventional to pick out an orthogonal basis $\{T^a\}$ for the Cartan subalgebra:
\begin{equation}
		b(T^a, T^b)=T(F)\delta^{ab},
		\label{eq:Ortho-Lie-basis}
\end{equation}
where $T(F)$ is an arbitrary normalization. Using this basis we can expand any element of $\mathfrak h$  as:
\begin{equation}
		t=\frac{1}{T(F)}\sum_{i} b(t,T^a)T^a.
		\label{eq:h-ortho-exp}
\end{equation}
We can also use this basis to write any weight $w$ as an $l$ component vector $w^a=w(T^a)$. Using \eqref{eq:weights_inner_product} we can write $t_w$, the corresponding element of $\mathfrak h$ as:
\begin{equation}
		t_w=\frac{1}{T(F)}\sum_{a} w^a T^a\equiv \frac{1}{T(F)} w \cdot H,
		\label{eq:h^*-ortho-exp}
\end{equation}
and we have:
\begin{equation}
		w\cdot v=b(t_w,t_v)=\frac 1{T(F)}\sum_i w_i v_i.
		\label{eq:weight-length}
\end{equation}
In this notation the length of the roots/weights as well as their inner products do not depend\footnote{Equivalently, we can define the inner product as $w\cdot v=\sum_i w_i v_i$, in which case we would have to take $T(F)=1$ to obtain the same results.} on the normalization $T(F)$.

For completeness, a summary of the Lie algebra data used in section \ref{sec:conf-bound} is given in table \ref{tab:Lie-data}.

\TABLE{
\begin{tabular}{|c|c|c|c|c|c|}
	\hline
		Group				&			Algebra		&			Root lengths				&			Kac labels					&	Rep'n		&	$T(R)$	\\\hline
	\multirow{3}{*}{$SO(2N+1)$}	&	\multirow{3}{*}{$B_N$}	&	\multirow{3}{*}{$\left\{2,\ldots,2,1\right\}$}	&	\multirow{3}{*}{$\left\{1,2,\ldots,2,1\right\}$}	&	fund.			&	$1$	\\  \cline{5-6}
							&						&									&										&	sym.			&	2N+3\\		\cline{5-6}
							&						&									&										&	adj.			&	2N-1\\\hline
	\multirow{3}{*}{$SP(2N)$}		&	\multirow{3}{*}{$C_N$}	&	\multirow{3}{*}{$\left\{1,\ldots,1,2\right\}$}	&	\multirow{3}{*}{$\left\{1,\ldots,1\right\}$}		&	fund.			&	$0.5$	\\  \cline{5-6}
							&						&									&										&	anti-sym.		&	N-1\\		\cline{5-6}
							&						&									&										&	adj.			&	N+1\\\hline
	\multirow{3}{*}{$SO(2N)$}		&	\multirow{3}{*}{$D_N$}	&	\multirow{3}{*}{$\left\{2,\ldots,2\right\}$}	&	\multirow{3}{*}{$\left\{1,2,\ldots,2,1,1\right\}$}	&	fund.			&	$1$	\\  \cline{5-6}
							&						&									&										&	sym.			&	2N+2\\		\cline{5-6}
							&						&									&										&	adj.			&	2N-2\\\hline
\end{tabular}
\caption{Lie algebra data. The length of $\tilde\alpha$ is always 2 and its Kac label is 1 and is not included in the table.}
\label{tab:Lie-data}
} 

\bibliographystyle{JHEP}
\bibliography{biblio}

\end{document}